\pdfoutput=1
\documentclass[conference]{IEEEtran}

\usepackage{amsmath,amssymb,amsfonts}
\usepackage{graphicx}
\usepackage{booktabs}
\usepackage{pifont}
\usepackage{url}
\usepackage{tikz}
\usetikzlibrary{arrows.meta,positioning,fit,backgrounds,calc,shapes.geometric,shapes.symbols,shadows}
\usepackage[ruled,vlined,linesnumbered]{algorithm2e}
\usepackage{cite}

\newtheorem{proposition}{Proposition}
\newcommand{\headAcc}{72.0}

\newcommand{\headAccFLTrust}{69.2}

\newcommand{\headAccKrumStd}{15.8}
\newcommand{\headFPRFLTrust}{0.44}

\newcommand{\headRecallLFighter}{0.46}
\newcommand{\headRecallFLAME}{0.42}
\newcommand{\headASR}{3.2}
\newcommand{\headASRFedAvg}{32.8}
\newcommand{\headDet}{15}
\newcommand{\defMs}{30}
\newcommand{\defMsFedAvg}{34}

\newcommand{\evalOverhead}{6.8}
\newcommand{\cleanHonRaw}{18.3}

\newcommand{\cleanHonBudget}{8.0}

\newcommand{\cleanAccRaw}{16.3}
\newcommand{\cleanAcc}{54.3}

\newcommand{\cleanAccMid}{66.5}
\newcommand{\cleanAccMidRaw}{40.7}
\newcommand{\cleanHonMidRaw}{16.0}
\newcommand{\skewAcc}{49.1}

\newcommand{\skewAccFedAvg}{43.8}

\newcommand{\skewAccLFighter}{41.6}
\newcommand{\skewAccFLTrust}{39.3}

\newcommand{\skewFPRRaw}{0.94}
\newcommand{\skewRecall}{0.96}

\newcommand{\fakeRecallRaw}{0.67}
\newcommand{\fakeRecallCROnly}{0.62}

\newcommand{\ratioAccLow}{74.1}
\newcommand{\ratioAccFedAvgLow}{74.3}
\newcommand{\ratioAccThirty}{72.4}
\newcommand{\ratioAccFedAvgThirty}{61.0}
\newcommand{\ratioAccLFighterThirty}{64.9}
\newcommand{\ratioAccFLTrustThirty}{70.8}
\newcommand{\bnsafeHead}{57.8}
\newcommand{\mnistAcc}{98.7}

\newcommand{\skewFull}{2/3}
\newcommand{\skewDetHits}{70}
\newcommand{\skewMissed}{1}
\newcommand{\fltrustNanRuns}{3}
\newcommand{\fltrustNanRounds}{6}

\newcommand{\cleanAccStd}{9.8}

\newcommand{\ccOverrideShare}{0.6}
\newcommand{\ratioFPRmax}{0.14}
\newcommand{\matchedOurs}{173}
\newcommand{\matchedFedAvg}{183}
\newcommand{\matchedFLTrust}{1269}
\newcommand{\matchedKrum}{4109}
\newcommand{\matchedFLAME}{3465}

\newcommand{\nRuns}{471}
\newcommand{\ccHeadAcc}{72.2}

\newcommand{\ccSkewRecall}{0.04}

\newcommand{\fsFound}{1}
\newcommand{\fsTotal}{24}
\newcommand{\fedWCleanAcc}{59.7}
\newcommand{\fedWCleanStd}{10.5}

\newcommand{\fedWSkewAcc}{44.8}

\newcommand{\fedWSkewASR}{21.4}
\newcommand{\pfAccFedAvgHalf}{71.9}
\newcommand{\pfAccFedAvgQuarter}{73.6}
\newcommand{\fedCleanAcc}{57.6}
\newcommand{\fedCleanStd}{9.4}

\newcommand{\pfAccHalf}{71.3}
\newcommand{\pfAccQuarter}{72.1}
\newcommand{\pfDetHalf}{24}
\newcommand{\pfDetQuarter}{39}
\newcommand{\fakeDet}{13}
\newcommand{\fsAcc}{43.1}
\newcommand{\fsRecall}{0.04}

\newcommand{\sizeLowMin}{143}
\newcommand{\sizeLowMax}{10{,}704}
\newcommand{\sizeHighMin}{1{,}244}
\newcommand{\sizeHighMax}{4{,}835}
\newcommand{\aucHighMin}{0.73}
\newcommand{\aucHighMax}{0.91}
\newcommand{\aucLowMin}{0.56}
\newcommand{\aucLowMax}{0.65}
\newcommand{\trajSeed}{19}
\newcommand{\fsAccFedAvg}{43.8}

\newcommand{\skewWinsFedW}{2}
\newcommand{\skewWinsFed}{3}
\newcommand{\skewASR}{7.5}
\newcommand{\skewASRFedAvg}{18.7}
\newcommand{\skewASRBest}{3}

\newcommand{\fakeAucResHighMin}{0.70}
\newcommand{\fakeAucResHighMax}{0.73}
\newcommand{\fakeAucRawLowMin}{0.64}
\newcommand{\fakeAucRawLowMax}{0.73}
\newcommand{\fakeAucResLowMin}{0.46}
\newcommand{\fakeAucResLowMax}{0.58}
\newcommand{\sensAccMin}{98.4}
\newcommand{\sensAccMax}{98.8}
\newcommand{\ppAcc}{71.0}
\newcommand{\ppAccStd}{1.0}

\newcommand{\ppAccFedAvg}{46.1}
\newcommand{\ppAccLFighter}{63.6}
\newcommand{\ppAccLFighterStd}{6.5}
\newcommand{\ppAccFLTrust}{62.1}
\newcommand{\ppASR}{3.3}
\newcommand{\ppASRFedAvg}{30.6}
\newcommand{\ppASRLFighter}{10.0}
\newcommand{\ppASRFLTrust}{11.1}

\newcommand{\ppFull}{3/3}
\newcommand{\ppDet}{34}

\newcommand{\pfRecLFighterHalf}{0.46}
\newcommand{\pfRecLFighterQuarter}{0.17}
\newcommand{\pfRecFLTrustHalf}{0.67}
\newcommand{\pfRecFLTrustQuarter}{0.50}

\newcommand{\sensFPRmax}{0.09}
\newcommand{\sensRecallMin}{1.00}

\begin{document}

\title{\fontsize{21}{25}\selectfont AntiFLipper: A Secure and Efficient Defense Against\\ Label-Flipping Attacks in Federated Learning}

\author{\IEEEauthorblockN{Aashnan Rahman$^{1}$, Abid Hasan$^{2}$, Sherajul Arifin$^{3}$, Faisal Haque Bappy$^{4}$, Tahrim Hossain$^{5}$,\\ Tariqul Islam$^{6}$, Abu Raihan Mostofa Kamal$^{7}$, and Md. Azam Hossain$^{8}$}
\IEEEauthorblockA{
$^{1, 2, 7, 8}$ Islamic University of Technology, Bangladesh.\\
$^{3}$ United International University, Bangladesh.\\
$^{4, 5, 6}$ University of Maryland, Baltimore County, United States.}
}

\maketitle

\begin{abstract}
In federated learning, malicious clients can corrupt a shared classifier by flipping the labels of their local training data. A natural way to detect them is to ask each client how accurately the global model predicts its local labels. However, because the global model is more accurate on some classes than others, honest clients with different class proportions can report very different accuracies. Malicious clients can also falsify their reports. We present AntiFLipper, a lightweight defense that needs no clean validation data at the server. Each client reports its per-class sample and correct-prediction counts, from which the server estimates the global model's accuracy on every class. The server then compares each client's reported accuracy with the accuracy expected for its class proportions, and a check on the direction of each client's final-layer update provides a second detection signal. On MNIST and CIFAR-10, the 84-byte report adds less than 0.1\% to each model upload, and server processing time is comparable to FedAvg in a matched benchmark. On CIFAR-10 with moderate label skew and 40\% of clients flipping all their labels, AntiFLipper reaches \headAcc\% mean test accuracy, versus \headAccFLTrust\% for the strongest of six baselines, and detects every attacker without excluding any honest client. In the same setting, detection remains complete in separate tests with falsified reports, partial label flipping and 50\% client participation. Under severe label skew without attackers, AntiFLipper reduces wrongful exclusions of honest clients from \cleanHonRaw{} of 20 on average with unadjusted accuracy feedback to zero.
\end{abstract}


\section{Introduction}
Federated learning (FL) lets clients train a shared model while keeping their training data local~\cite{fedavg-pmlr-v54-mcmahan17a,kairouz2021advances}. In a \emph{label-flipping attack}, a malicious client deliberately flips the labels of local training examples and trains on the altered data. It sends its update through the usual protocol. Aggregating those updates can reduce the global model's accuracy on the classes targeted by the attack~\cite{biggio2011support,tolpegin2020data}.

Some defenses use a small, correctly labeled reference dataset at the server to judge which updates to trust~\cite{cao2020fltrust}. Such a dataset can be hard to obtain when labeled examples must stay with the clients. Client-side accuracy feedback offers an alternative. Each client evaluates the global model on its local data and reports the accuracy~\cite{andreina2021baffle,jiang2023dataquality}. When labels are flipped, predictions that match the original labels count as errors. A low score is ambiguous, however. An honest client may also report low accuracy if its dataset contains more examples from classes the model predicts less reliably. The server must also account for clients that falsify their reports.

\textbf{AntiFLipper} uses per-class counts from participating clients to estimate how accurate the global model is on each class. It uses those estimates and the mix of labels at each client to calculate that client's expected accuracy. AntiFLipper subtracts this expectation from the accuracy the client reports. This difference is the \emph{class-conditional residual}. A negative residual means the client's reported accuracy is below what its label mix would suggest. The server also compares the direction of each client's final-layer update with a reference computed from the participating clients' updates. It combines the two signals across rounds to update client trust and decide which clients to exclude.

AntiFLipper makes the following contributions.
\begin{itemize}
\item \textbf{Mitigation and client-level detection.} With 40\% of clients flipping all their labels, AntiFLipper reaches \headAcc\% mean accuracy on moderately skewed CIFAR-10, versus \headAccFLTrust\% for the strongest baseline, and detects all attackers with no false positives. It is the only tested method that does so on both CIFAR-10 and MNIST, with IID and non-IID data, and on MNIST it matches the best accuracy (\mnistAcc\%).
\item \textbf{Low overhead.} AntiFLipper requires no server-held validation data, adds 84 bytes per participating client per round, and has server processing time comparable to FedAvg in a matched benchmark.
\item \textbf{Detection across tested conditions.} Under moderate skew, AntiFLipper detects all attackers at tested shares of 10--40\% and in separate tests of partial flips, falsified reports and 50\% participation. In clean, severely skewed CIFAR-10 runs, it excludes no honest clients, compared with \cleanHonRaw{} of 20 on average for raw accuracy feedback.
\end{itemize}

Detection remains limited when attackers falsify their reports under severe label skew, as shown in Section~\ref{sec:results}.

The following sections review related work (Section~\ref{sec:related}), define the system and threat model (Section~\ref{sec:model}), and present AntiFLipper (Section~\ref{sec:design}). Sections~\ref{sec:setup} and~\ref{sec:results} describe the experiments and results, and Section~\ref{sec:analysis} analyzes the method. Sections~\ref{sec:limits} and~\ref{sec:conclusion} discuss limitations and conclude the paper.

\section{Related Work}
\label{sec:related}
Early work studied label flipping as deliberate label noise in centralized training~\cite{biggio2011support,munoz2017towards}, and certified defenses bound the effects of poisoned samples for convex learners~\cite{steinhardt2017certified}. In federated learning, Tolpegin et al.~\cite{tolpegin2020data} showed that label flipping can reduce accuracy on the affected classes even with a minority of attackers. Its impact depends on how data are distributed~\cite{lavaur2024systematic} and can be limited when very few clients are compromised~\cite{shejwalkar2022backdrawingboard}. We study settings with up to 40\% malicious clients. Label flipping can occur during ordinary local training, so defenses developed for crafted updates or backdoors address a different threat~\cite{bhagoji2019analyzing,fang2020local,shejwalkar2021manipulating,bagdasaryan2020backdoor,wang2020attack,sun2019really}.

Robust aggregation limits the influence of suspicious updates. Coordinate-wise median and trimmed mean compute statistics across clients for every model parameter~\cite{yin2018byzantine}, and the geometric median requires repeated passes over the updates~\cite{pillutla2022robust}. Multi-Krum compares full updates in pairs, so its cost grows quadratically with the number of participating clients~\cite{NIPS2017_f4b9ec30}. Bulyan requires at least $4f+3$ updates for $f$ attackers~\cite{mhamdi2018hidden}, which exceeds the 20 clients available in our CIFAR-10 runs with eight attackers. Centered clipping limits the magnitude of unusually large moves~\cite{karimireddy2021history}, but label flipping can produce ordinary-sized updates. Non-IID data can also pull honest updates in different directions, weakening proximity as a signal~\cite{zhao2018noniid,li2020federated,karimireddy2022bucketing}.

Detection methods add other forms of server work. FLAME clusters models before clipping and adding noise~\cite{nguyen2022flame}, and FLDetector uses each client's update history to predict its next update~\cite{zhang2022fldetector}. FoolsGold also keeps past updates, using their similarity to identify coordinated clients~\cite{fung2020limitations}. Tolpegin et al.~\cite{tolpegin2020data} and LFighter~\cite{jebreel2024lfighter} reduce the dimension of the clustering problem by focusing on the output layer. Their decisions still depend on whether honest and malicious updates separate in that space. FLTrust trains a reference model at the server each round~\cite{cao2020fltrust}, and HSCS evaluates models on a clean server-held dataset to score clients by class~\cite{li2023hscs}. Both require labeled examples at the server. HSCS also adds model evaluations, and FLTrust adds a training step.

Client feedback uses examples already available to the participants. BaFFLe asks validation clients to evaluate the global model and vote~\cite{andreina2021baffle}, adding client computation and communication. FedAlign extracts global and class-level statistics with a variational autoencoder and uses them to adjust trust during training~\cite{fedalign2026}. LFR-PPFL protects its similarity calculations with homomorphic encryption~\cite{shen2024lfrppfl}, which adds cryptographic work. FedCorr repairs client labels under heterogeneous noise~\cite{xu2022fedcorr}, a different goal from identifying clients that deliberately alter labels and may falsify reports. AntiFLipper uses per-class counts to interpret each client's accuracy in the context of its label mix and adds a check on the direction of the final-layer update. The defense needs a local evaluation pass and a small report from each participating client. Its trust decision does not involve clustering full updates or training a reference model at the server.

\section{System and Threat Model}
\label{sec:model}
A server coordinates $N$ clients over $T$ rounds. Client $j$ holds a dataset $D_j$ with $K$ possible classes. Its label mix is $\pi_{jk}=n_{jk}/|D_j|$, where $n_{jk}$ is the number of its examples with label $k$. In round $t$, the server sends the global model $M_t$ to the sampled clients $S_t$. By default, every client that has not been excluded takes part. Each sampled client trains $M_t$ for $E$ local epochs and returns the resulting model $m_{j,t}$ along with a report. The report contains the accuracy $a_j$ of $M_t$ on its local labels, its per-class correct counts $\mathbf{c}_j$, and its per-class sample counts $\mathbf{n}_j$. The server tracks a trust score $\tau_j$ for each client, with the scores summing to one over the active clients, and records excluded clients in $\beta$.

The adversary controls a fraction $p\le 0.4$ of the clients. Malicious clients use a fixed label permutation $\sigma$ that swaps $0$ with $1$, $2$ with $3$, and so on through $8$ and $9$. They train on the relabeled data using the usual training code. A full flip relabels every local example in each round. A partial flip relabels a fixed fraction of local examples, with a new random subset drawn each round.

We test two reporting behaviors. Under honest reporting, attackers evaluate $M_t$ on the labels they use for training. Under report faking, they evaluate $M_t$ on their original labels and report those statistics. The threat model excludes crafted updates such as gradient scaling and backdoor attacks~\cite{bagdasaryan2020backdoor,fang2020local}.

\section{Methodology}
\label{sec:design}
AntiFLipper compares a client's reported accuracy with the accuracy expected for its label mix and checks its last-layer update against those of its peers. These two signals determine the weight given to each update and whether a client qualifies for exclusion. Fig.~\ref{fig:arch} shows one round, and Algorithm~\ref{alg:af} shows how the server uses client reports and model updates to detect suspicious clients and aggregate the retained updates.

\begin{figure}[t]
\centering
\resizebox{\columnwidth}{!}{
\definecolor{repc}{RGB}{38,99,170}\definecolor{repf}{RGB}{226,236,250}
\definecolor{updc}{RGB}{204,110,20}\definecolor{updf}{RGB}{253,237,219}
\definecolor{okc}{RGB}{22,150,100}\definecolor{badc}{RGB}{214,48,49}
\definecolor{barf}{RGB}{222,222,222}
\begin{tikzpicture}[
  font=\sffamily\small, >={Stealth[length=5pt,width=4pt]}, line cap=round,
  bar/.style={rounded corners=5pt, fill=barf, minimum height=0.62cm, align=center, inner sep=2pt},
  tok/.style={circle, minimum size=0.62cm, inner sep=0pt, font=\sffamily\footnotesize\bfseries},
  dia/.style={diamond, aspect=1, minimum size=0.82cm, inner sep=0pt, rounded corners=3pt, font=\sffamily\footnotesize\bfseries},
  group/.style={draw=black!45, dotted, line width=0.7pt, rounded corners=6pt},
  lbl/.style={font=\sffamily\scriptsize, align=left, text=black!75, anchor=west},
  w/.style={line width=0.9pt},
]
\def\xa{1.0}\def\xb{2.5}\def\xc{4.0}
\def\xl{4.6}   
\def\xm{7.65}  
\def\ybus{0.95}
\draw[dashed, line width=0.8pt, rounded corners=9pt, black!70] (0.15,1.3) rectangle (8.85,7.05);
\node[font=\sffamily\small\bfseries, fill=white, inner sep=3pt, anchor=west] at (0.4,7.05) {Server};
\foreach \x/\n/\name/\col in {\xa/a/Client 1/black, \xb/b/Client 2/black, \xc/c/Attacker/badc}{
  \node[draw=\col, line width=1.1pt, rounded corners=4pt, minimum width=1.32cm, minimum height=0.95cm, fill=white] (cl\n) at (\x,0) {};
  \begin{scope}[shift={(\x-0.36,0.04)}, \col, line width=0.7pt]
    \draw[fill=white] (-0.16,0.16) -- (-0.16,-0.14) arc[start angle=180, end angle=360, x radius=0.16, y radius=0.055] -- (0.16,0.16);
    \draw[fill=white] (0,0.16) ellipse[x radius=0.16, y radius=0.055];
    \draw[line width=0.5pt] (-0.16,0.01) arc[start angle=180, end angle=360, x radius=0.16, y radius=0.055];
  \end{scope}
  \node[font=\sffamily\scriptsize, text=\col] at (\x+0.2,0.06) {$m_j$};
  \node[font=\sffamily\scriptsize, text=\col] at (\x,-0.33) {\name};
}
\node[font=\sffamily\scriptsize, text=badc, align=center, anchor=west] at (\xc+0.72,0.0) {flips\\labels};
\node[group, minimum width=3.9cm, minimum height=0.95cm] at (\xb,2.05) {};
\foreach \x/\n/\c in {\xa/a/repc, \xb/b/repc, \xc/c/badc}
  \node[tok, fill=repf, draw=\c, line width=1.1pt] (r\n) at (\x,2.05) {M1};
\node[lbl] at (\xl,2.05) {Per-class\\reports\\($4{+}8K$\,B)};
\node[bar, minimum width=4.95cm] (m2) at (2.925,3.1) {\textbf{M2:} Class-conditional residual \ $r_j$};
\node[group, minimum width=3.9cm, minimum height=1.15cm] at (\xb,4.15) {};
\node[dia, fill=okc, text=white] (ta) at (\xa,4.15) {M3};
\node[dia, fill=okc, text=white] (tb) at (\xb,4.15) {M3};
\node[dia, fill=white, draw=badc, line width=1.1pt, text=badc] (tc) at (\xc,4.15) {M3};
\node[lbl] at (\xl,4.15) {Trust\\update $\tau_j$};
\node[bar, minimum width=4.95cm] (m5) at (2.925,5.15) {\textbf{M5:} Flags \& capped exclusion};
\node[bar, minimum width=8.1cm]  (m6) at (4.5,6.2)  {\textbf{M6:} Trust-weighted aggregation over $S_t\setminus\beta$};
\node[rounded corners=5pt, fill=updf, draw=updc, line width=0.9pt, minimum width=1.8cm, minimum height=3.9cm] (m4) at (\xm,3.55) {};
\node[align=center, anchor=north, inner sep=0pt] at ($(m4.north)+(0,-0.18)$) {\textbf{M4:}\\Update\\consistency\\check};
\draw[updc!60, line width=0.5pt] ($(m4.center)+(-0.6,0.1)$) -- ($(m4.center)+(0.6,0.1)$);
\node[font=\footnotesize] at ($(m4.center)+(0,-0.5)$) {$\cos(\delta_j,\tilde\delta)$};
\node[font=\sffamily\scriptsize, align=center, text=black!75] at ($(m4.center)+(0,-1.3)$) {last layer\\vs.\ median};
\foreach \n in {a,b,c}
  \draw[w, updc] ($(cl\n.north)+(0.42,0)$) -- ($(cl\n.north)+(0.42,\ybus-0.475)$);
\draw[w, updc, ->] ($(\xa+0.42,\ybus)$) -- (\xm,\ybus) -- (m4.south);
\node[font=\sffamily\scriptsize, text=updc, anchor=north] at (6.45,\ybus-0.04) {updates $m_j$};
\foreach \n/\c in {a/repc, b/repc, c/badc}
  \draw[w, \c, ->, preaction={draw=white, line width=3.4pt, shorten <=2pt, shorten >=2pt}] (cl\n.north) -- (r\n.south);
\foreach \n/\c in {a/black, b/black, c/badc}{
  \draw[w, \c, ->] (r\n.north) -- (r\n.north |- m2.south);
  \draw[w, \c, ->] (t\n.south |- m2.north) -- (t\n.south);
  \draw[w, \c, ->] (t\n.north) -- (t\n.north |- m5.south);
}
\foreach \x in {\xa,\xb}
  \draw[w, ->] (\x,5.46) -- (\x,5.89);
\draw[w, badc, dashed] (\xc,5.46) -- (\xc,5.6);
\node[circle, fill=badc, text=white, inner sep=0.6pt, font=\sffamily\tiny\bfseries] at (\xc,5.69) {$\times$};
\node[font=\sffamily\scriptsize, text=badc, anchor=west] at (\xc+0.18,5.69) {excluded};
\draw[w, updc, ->] (m4.west |- m5) -- node[above, font=\sffamily\scriptsize, text=updc] {flag} (m5.east);
\draw[w, updc, ->] (m4.west |- m2) -- node[above, font=\sffamily\scriptsize, text=updc] {$r_j{\leftarrow}\min$} (m2.east);
\draw[w, updc, ->] (m4.north) -- (m4.north |- m6.south);
\node[cloud, cloud puffs=11, cloud puff arc=110, aspect=2.2, draw=black!75, line width=0.9pt, fill=white,
      minimum width=2.3cm, minimum height=0.95cm, font=\sffamily\footnotesize, align=center] (gm) at (4.5,7.95) {Global model\\$M_{t+1}$};
\draw[w, ->] (m6.north) -- (gm.south);
\draw[w, ->] (gm.west) -- (-0.45,7.95) -- (-0.45,-0.95) -- (\xc,-0.95) -- (clc.south);
\foreach \n in {a,b} \draw[w, ->] (cl\n.south |- 0,-0.95) -- (cl\n.south);
\node[font=\sffamily\scriptsize, rotate=90, anchor=south] at (-0.45,3.4) {broadcast to sampled clients};
\end{tikzpicture}}
\caption{One round of AntiFLipper. Clients send reports (blue) and model updates (orange). An inconsistent update sets the client's residual to the round minimum and earns a flag after warm-up. In this example, the attacker (red) reaches the flag limit and is excluded within the cap. The server weights the retained updates by trust.}
\label{fig:arch}
\end{figure}

\textbf{M1: Client report.} Before local training, client $j$ evaluates the received model $M_t$ in one pass over its local data. It reports overall accuracy $a_j$ and, for each class $k$, the correct count $c_{jk}$ and sample count $n_{jk}$. The counts determine $a_j$, which is included explicitly for the accuracy-only variants. The 32-bit fields total $4+8K$ bytes per report, or 84 bytes for $K{=}10$.

\textbf{M2: Class-conditional residual.} The server pools per-class accuracy over the sampled clients, weighted by trust. Weighting these class accuracies by the label proportions in client $j$'s report gives the accuracy expected for that client, and the residual $r_j$ is its reported accuracy minus this expectation.
\begin{equation}
\hat A_k=\frac{\sum_{i\in S_t} \tau_i c_{ik}}{\sum_{i\in S_t} \tau_i n_{ik}},\qquad
r_j=a_j-\sum_k \pi_{jk}\hat A_k .
\label{eq:cr}
\end{equation}
The expectation uses only classes in the client's report with a defined pooled accuracy. The server renormalizes the label proportions over those classes and sets $r_j=0$ if none remain. An honest client can report low accuracy because it holds classes that the model often misclassifies. Its residual stays near zero if the pooled accuracies reflect the same pattern. For an attacker reporting on flipped labels, the residual can be negative when the global model predicts the original labels reliably. When attackers supply much of the reference for the classes they report, their own reports can lower the expected accuracy and shrink this gap (Section~\ref{sec:analysis}).

\textbf{M3: Trust update.} The server first applies the consistency check in M4. For each sampled client, it then adjusts trust by the signed squared difference between $r_j$ and the mean residual $\bar r$ of the sampled clients.
\begin{equation}
\tau_j \leftarrow \max\!\big(0,\ \tau_j + \operatorname{sgn}(r_j-\bar r)\,\eta\,(r_j-\bar r)^2\big).
\label{eq:trust}
\end{equation}
The server then normalizes trust over all active clients. The squared term gives more weight to large residual gaps. Trust persists across rounds, also for clients that are not sampled, so evidence accumulates over every round in which a client takes part.

\textbf{M4: Update-consistency check.} Let $\delta_j$ be client $j$'s change in the classification layer (weights and bias), and let $\tilde\delta$ be the coordinate-wise median of the received changes. The server compares each cosine similarity $s_j=\cos(\delta_j,\tilde\delta)$ with the median similarity using $(s_j-\operatorname{med}(s))/(1.4826\,\mathrm{MAD}(s)+\epsilon)$, where MAD is the median absolute deviation and $\epsilon=10^{-6}$. With at least three updates, a client scoring below $-z$ is \emph{inconsistent}, and its residual is replaced by the round minimum before~(\ref{eq:trust}).

\textbf{M5: Flags and capped exclusion.} After $g$ warm-up rounds, a sampled client receives one flag if $n_t\tau_j<\rho$ or M4 marks its update inconsistent. Here $n_t$ counts all active clients, so the trust threshold is a fraction $\rho$ of the uniform weight $1/n_t$. Flags accumulate across rounds. A sampled client qualifies for exclusion at $f_{\max}$ flags in total. The server excludes at most $b=\lfloor0.4N\rfloor$ clients overall, prioritizing those with the lowest trust when more qualify than the remaining budget allows.

\textbf{M6: Aggregation.} The server averages the models of the sampled clients that are not excluded, weighted by trust, so that $M_{t+1}=\sum_{j\in S_t\setminus\beta}\tau_j\,m_{j,t}\big/\sum_{j\in S_t\setminus\beta}\tau_j$. Equal trust among retained clients gives equal-weight FedAvg. The two checks use class counts and last-layer updates and require no server-side validation data.

\setlength{\textfloatsep}{8pt}
\begin{algorithm}[t]
\caption{AntiFLipper: server side of round $t$}
\label{alg:af}
\small
\KwIn{models $m_{j,t}$, reports $(a_j,\mathbf{c}_j,\mathbf{n}_j)$ for $j\in S_t$, trust $\tau$, flag counts $f_j$, excluded set $\beta$}
compute $\hat A_k$ over $S_t$ by~(\ref{eq:cr}) for classes with positive pooled counts\;
\lForEach{$j\in S_t$}{compute $r_j$ by~(\ref{eq:cr}), with the M2 renormalization}
$I\leftarrow\varnothing$\;
\If{$|S_t|\ge 3$}{
 $\tilde\delta\leftarrow$ coordinate-wise median of last-layer updates $\{\delta_j\}$\;
 $s_j\leftarrow\cos(\delta_j,\tilde\delta)$ for every $j\in S_t$\;
 $I\leftarrow\{j:\ (s_j-\operatorname{med}(s))/(1.4826\,\mathrm{MAD}(s)+\epsilon)<-z\}$\;
 \lForEach{$j\in I$}{$r_j\leftarrow\min_{i\in S_t} r_i$}
}
update $\tau$ with~(\ref{eq:trust}) and normalize over $j\notin\beta$\;
\If{$t\ge g$}{
 \ForEach{$j\in S_t$}{\lIf{$n_t\tau_j<\rho$ \textbf{\upshape or} $j\in I$}{$f_j\leftarrow f_j+1$}}
 $Q\leftarrow\{j\in S_t:\ f_j\ge f_{\max}\}$\;
 \lIf{$|Q|>b-|\beta|$}{keep the $b-|\beta|$ clients of $Q$ with lowest trust}
 add $Q$ to $\beta$ and normalize $\tau$\;
}
\Return $M_{t+1}=\sum_{j\in S_t\setminus\beta}\tau_j m_{j,t}/\sum_{j\in S_t\setminus\beta}\tau_j$
\end{algorithm}

\section{Experimental Setup}
\label{sec:setup}
\textbf{Data and models.} CIFAR-10~\cite{CIFAR-krizhevsky2009learning} is trained with ResNet-18~\cite{resnet-he2016deep} (11.2\,M parameters) over 20 clients and 100 rounds, and MNIST~\cite{MNIST-lecun1998gradient} with a CNN of 21.8\,k parameters over 100 clients and 200 rounds. IID and non-IID splits use the same number of clients. Non-IID splits divide the samples of each class among the clients in proportions drawn from $\mathrm{Dir}(\alpha_D)$, with $\alpha_D{=}1$ (moderate skew) unless stated otherwise and $\alpha_D{=}0.1$ for strong skew. Client datasets are therefore unequal in size. On CIFAR-10 they range from \sizeHighMin{} to \sizeHighMax{} samples at $\alpha_D{=}1$ and from \sizeLowMin{} to \sizeLowMax{} at $\alpha_D{=}0.1$. Unless stated otherwise, all clients take part in every round and train for $E{=}3$ epochs with SGD (learning rate 0.01, momentum 0.9, weight decay $5\cdot10^{-4}$, batch size 32 for CIFAR-10 and 64 for MNIST).

\textbf{Attacks.} By default 40\% of the clients perform a full flip and report honestly. This is the largest share our threat model allows and the hardest case, where undefended training and several baselines break down (Fig.~\ref{fig:ratio}). We also evaluate partial flips of 50\% and 25\%, partial participation (a random half of the active clients in every round), report faking at $\alpha_D{=}1$ and $0.1$, shares of 10--30\%, and runs without attackers, in which every exclusion is a false positive.

\textbf{Baselines.} We compare with FedAvg, coordinate-wise Median, Multi-Krum ($f=\lfloor pN\rfloor$), FLAME, LFighter and FLTrust. FLTrust additionally receives a class-balanced root set of 100 clean samples. Its reference implementation rescales all tensors including BatchNorm statistics, so we also report a variant that rescales only trainable parameters. Bulyan is not applicable, as $N{=}20$ and $f{=}8$ violate $N\ge4f+3$. FedAvg denotes equal client weights, matching AntiFLipper's initial trust. We also run the original sample-weighted FedAvg~\cite{fedavg-pmlr-v54-mcmahan17a} under strong skew and, for reference, in the ablation. Later experiments use LFighter, designed for label flipping, and FLTrust, the strongest baseline in the main setting, in its BN-safe variant for partial flips and participation. Ablations replace the residual with the raw report $a_j$, remove the consistency check (CC) or the cap, or use CC alone with uniform trust. All methods share partitions and attacker sets.

\textbf{Parameters and metrics.} AntiFLipper uses $\eta{=}0.1$, $\rho{=}0.05$, $f_{\max}{=}6$, $g{=}2$, $z{=}3$ and $b{=}0.4N$. Results are means and standard deviations over three seeds, \nRuns{} training runs in total, and the sensitivity study uses separate held-out seeds. Accuracy and attack success rate (ASR, the share of test predictions equal to $\sigma(y)$) are averaged over the last 10 rounds, since single rounds fluctuate by several points under strong skew. Recall (Rec.) and false-positive rate (FPR) refer to the clients excluded, or given zero weight by baselines, in the final round, and are marked n/a for methods that exclude no client. Tables give accuracy and ASR in percent, with the standard deviation of accuracy over seeds in small type. Server time covers all checks, trust updates and aggregation. We measure it during training, excluding rounds that overlapped other workloads, and in a matched benchmark that gives every method the same 20 updates.

\section{Results}
\label{sec:results}
\begin{table*}[t]
\centering
\caption{Comparison with existing defenses (40\% attackers, three seeds).}
\label{tab:main}
\footnotesize
\setlength{\tabcolsep}{2.2pt}
\begin{tabular}{@{}l cccc cccc cccc cccc@{}}
\toprule
& \multicolumn{4}{c}{\textbf{CIFAR-10, non-IID}} & \multicolumn{4}{c}{\textbf{CIFAR-10, IID}} & \multicolumn{4}{c}{\textbf{MNIST, non-IID}} & \multicolumn{4}{c}{\textbf{MNIST, IID}} \\
\cmidrule(lr){2-5}\cmidrule(lr){6-9}\cmidrule(lr){10-13}\cmidrule(l){14-17}
\textbf{Method} & Acc.\,$\uparrow$ & ASR\,$\downarrow$ & Rec.\,$\uparrow$ & FPR\,$\downarrow$ & Acc.\,$\uparrow$ & ASR\,$\downarrow$ & Rec.\,$\uparrow$ & FPR\,$\downarrow$ & Acc.\,$\uparrow$ & ASR\,$\downarrow$ & Rec.\,$\uparrow$ & FPR\,$\downarrow$ & Acc.\,$\uparrow$ & ASR\,$\downarrow$ & Rec.\,$\uparrow$ & FPR\,$\downarrow$ \\
\midrule
FedAvg & 44.2{\scriptsize$\pm$0.5} & 32.8 & n/a & n/a & 42.1{\scriptsize$\pm$0.1} & 36.4 & n/a & n/a & 49.8{\scriptsize$\pm$0.3} & 48.9 & n/a & n/a & 49.6{\scriptsize$\pm$0.1} & 49.2 & n/a & n/a \\
Median & 57.9{\scriptsize$\pm$4.2} & 16.7 & n/a & n/a & 62.9{\scriptsize$\pm$1.4} & 13.1 & n/a & n/a & 97.6{\scriptsize$\pm$0.0} & 0.4 & n/a & n/a & 97.5{\scriptsize$\pm$0.2} & 0.3 & n/a & n/a \\
Multi-Krum & 60.8{\scriptsize$\pm$15.8} & 12.5 & 0.75 & 0.33 & 72.9{\scriptsize$\pm$0.3} & \textbf{3.1} & \textbf{1.00} & 0.17 & \textbf{98.7}{\scriptsize$\pm$0.2} & \textbf{0.1} & \textbf{1.00} & 0.03 & \textbf{98.7}{\scriptsize$\pm$0.1} & \textbf{0.1} & \textbf{1.00} & 0.03 \\
FLAME & 56.9{\scriptsize$\pm$7.7} & 17.7 & 0.42 & 0.17 & 71.1{\scriptsize$\pm$0.4} & 3.3 & \textbf{1.00} & 0.08 & \textbf{98.7}{\scriptsize$\pm$0.1} & \textbf{0.1} & \textbf{1.00} & 0.14 & \textbf{98.7}{\scriptsize$\pm$0.1} & \textbf{0.1} & \textbf{1.00} & 0.11 \\
LFighter & 59.0{\scriptsize$\pm$13.3} & 15.2 & 0.46 & 0.08 & 73.2{\scriptsize$\pm$0.3} & \textbf{3.1} & \textbf{1.00} & \textbf{0.00} & 98.5{\scriptsize$\pm$0.2} & 0.2 & 0.85 & \textbf{0.00} & \textbf{98.7}{\scriptsize$\pm$0.1} & \textbf{0.1} & \textbf{1.00} & \textbf{0.00} \\
FLTrust (reference)$^{\dagger}$ & 69.2{\scriptsize$\pm$1.1} & 5.2 & 0.92 & 0.44 & 63.1{\scriptsize$\pm$0.3} & 13.2 & 0.79 & 0.50 & 97.2{\scriptsize$\pm$0.1} & 0.3 & 0.98 & 0.66 & 97.5{\scriptsize$\pm$0.1} & 0.2 & 0.99 & 0.50 \\
FLTrust (BN-safe)$^{\dagger}$ & 57.8{\scriptsize$\pm$1.2} & 15.1 & 0.29 & 0.33 & 63.9{\scriptsize$\pm$2.3} & 12.0 & 0.38 & 0.22 & 97.2{\scriptsize$\pm$0.1} & 0.3 & 0.98 & 0.66 & 97.5{\scriptsize$\pm$0.1} & 0.2 & 0.99 & 0.50 \\
\midrule
\textbf{AntiFLipper} & \textbf{72.0}{\scriptsize$\pm$0.6} & \textbf{3.2} & \textbf{1.00} & \textbf{0.00} & \textbf{73.3}{\scriptsize$\pm$0.4} & \textbf{3.1} & \textbf{1.00} & \textbf{0.00} & \textbf{98.7}{\scriptsize$\pm$0.1} & \textbf{0.1} & \textbf{1.00} & \textbf{0.00} & \textbf{98.7}{\scriptsize$\pm$0.2} & \textbf{0.1} & \textbf{1.00} & \textbf{0.00} \\
\bottomrule
\end{tabular}
\\[2pt]\parbox{\linewidth}{\scriptsize Non-IID: $\alpha_D{=}1$. Reference: rescales all tensors including BatchNorm statistics. BN-safe: rescales trainable parameters only, identical on MNIST, which has no BatchNorm. $^{\dagger}$Uses 100 clean root samples at the server.}
\end{table*}

\begin{table*}[t]
\centering
\caption{Ablation of AntiFLipper's components (CIFAR-10, three seeds).}
\label{tab:ablation}
\footnotesize
\setlength{\tabcolsep}{1.6pt}
\begin{tabular}{@{}l ccc cccc cccc cc cccc@{}}
\toprule
& & & & \multicolumn{4}{c}{\textbf{40\% attack, $\alpha_D{=}1$}} & \multicolumn{4}{c}{\textbf{40\% report faking, $\alpha_D{=}1$}} & \multicolumn{2}{c}{\textbf{No attack, $\alpha_D{=}0.1$}} & \multicolumn{4}{c}{\textbf{40\% attack, $\alpha_D{=}0.1$}} \\
\cmidrule(lr){5-8}\cmidrule(lr){9-12}\cmidrule(lr){13-14}\cmidrule(l){15-18}
\textbf{Variant} & CR & CC & Cap & Acc.\,$\uparrow$ & ASR\,$\downarrow$ & Rec.\,$\uparrow$ & FPR\,$\downarrow$ & Acc.\,$\uparrow$ & ASR\,$\downarrow$ & Rec.\,$\uparrow$ & FPR\,$\downarrow$ & Acc.\,$\uparrow$ & FPR\,$\downarrow$ & Acc.\,$\uparrow$ & ASR\,$\downarrow$ & Rec.\,$\uparrow$ & FPR\,$\downarrow$ \\
\midrule
FedAvg & \ding{55} & \ding{55} & \ding{55} & 44.2{\scriptsize$\pm$0.5} & 32.8 & n/a & n/a & 44.2{\scriptsize$\pm$0.5} & 32.8 & n/a & n/a & 57.6{\scriptsize$\pm$9.4} & n/a & 43.8{\scriptsize$\pm$9.1} & 18.7 & n/a & n/a \\
FedAvg (weighted) & \ding{55} & \ding{55} & \ding{55} & 44.3{\scriptsize$\pm$1.9} & 33.0 & n/a & n/a & 44.3{\scriptsize$\pm$1.9} & 33.0 & n/a & n/a & 59.7{\scriptsize$\pm$10.5} & n/a & 44.8{\scriptsize$\pm$9.2} & 21.4 & n/a & n/a \\
\midrule
Raw reports & \ding{55} & \ding{55} & \ding{55} & 71.0{\scriptsize$\pm$0.2} & \textbf{3.2} & \textbf{1.00} & 0.08 & 72.4{\scriptsize$\pm$0.2} & 3.5 & 0.67 & 0.08 & 16.3{\scriptsize$\pm$5.9} & 0.92 & 13.0{\scriptsize$\pm$2.6} & 15.8 & 0.88 & 0.94 \\
Raw reports + cap & \ding{55} & \ding{55} & \ding{51} & 71.1{\scriptsize$\pm$0.7} & \textbf{3.2} & \textbf{1.00} & \textbf{0.00} & 72.5{\scriptsize$\pm$0.2} & 3.6 & 0.71 & 0.08 & 16.8{\scriptsize$\pm$6.4} & 0.40 & 12.5{\scriptsize$\pm$1.7} & 15.2 & 0.38 & 0.42 \\
Raw reports + CC & \ding{55} & \ding{51} & \ding{55} & 70.4{\scriptsize$\pm$0.5} & 3.3 & \textbf{1.00} & 0.25 & 71.6{\scriptsize$\pm$0.5} & 3.4 & 0.88 & 0.17 & 31.2{\scriptsize$\pm$15.2} & 0.85 & 28.9{\scriptsize$\pm$3.9} & 9.5 & \textbf{0.96} & 0.86 \\
CC only$^{\ddagger}$ & \ding{55} & \ding{51} & \ding{51} & \textbf{72.2}{\scriptsize$\pm$0.5} & \textbf{3.2} & \textbf{1.00} & \textbf{0.00} & 72.2{\scriptsize$\pm$0.5} & \textbf{3.2} & \textbf{1.00} & \textbf{0.00} & 55.2{\scriptsize$\pm$7.8} & 0.02 & 44.6{\scriptsize$\pm$7.6} & 19.0 & 0.04 & 0.03 \\
CR + cap (no CC) & \ding{51} & \ding{55} & \ding{51} & 70.7{\scriptsize$\pm$1.1} & 3.4 & \textbf{1.00} & \textbf{0.00} & \textbf{73.5}{\scriptsize$\pm$0.8} & 3.3 & 0.62 & \textbf{0.00} & \textbf{60.2}{\scriptsize$\pm$7.5} & 0.02 & \textbf{49.1}{\scriptsize$\pm$12.0} & 8.2 & \textbf{0.96} & \textbf{0.00} \\
\midrule
\textbf{AntiFLipper} & \ding{51} & \ding{51} & \ding{51} & 72.0{\scriptsize$\pm$0.6} & \textbf{3.2} & \textbf{1.00} & \textbf{0.00} & 72.0{\scriptsize$\pm$0.3} & 3.3 & \textbf{1.00} & \textbf{0.00} & 54.3{\scriptsize$\pm$9.8} & \textbf{0.00} & \textbf{49.1}{\scriptsize$\pm$11.9} & \textbf{7.5} & \textbf{0.96} & \textbf{0.00} \\
\bottomrule
\end{tabular}
\\[2pt]\parbox{\linewidth}{\scriptsize Weighted: FedAvg weighted by client dataset size~\cite{fedavg-pmlr-v54-mcmahan17a}. Faking: attackers report statistics on their original labels. FedAvg rows: no defense, for reference. FedAvg and CC only ignore reports, so their faking columns repeat the 40\% attack runs. Without attackers, FPR is the share of the 20 honest clients excluded. CC: $z{=}2.5$ in ``Raw reports + CC'', $z{=}3$ elsewhere. $^{\ddagger}$Uniform trust.}
\end{table*}

\subsection{Comparison with existing defenses}
On CIFAR-10 non-IID with 40\% attackers (Table~\ref{tab:main}), AntiFLipper reaches \headAcc\% accuracy, the highest of all methods, and lowers the ASR from \headASRFedAvg\% under FedAvg to \headASR\%. In every seed it excludes all attackers, on average by round \headDet, and no honest client. The strongest baseline, the reference FLTrust, reaches \headAccFLTrust\% but needs server data, flags \headFPRFLTrust{} of the honest clients and produced non-finite losses in \fltrustNanRounds{} rounds across \fltrustNanRuns{} of its 3 runs. Its BatchNorm-safe variant avoids these failures but reaches only \bnsafeHead\%. LFighter and FLAME find \headRecallLFighter{} and \headRecallFLAME{} of the attackers, and their accuracy, like that of Multi-Krum ($\pm$\headAccKrumStd), varies strongly across seeds. On CIFAR-10 IID, AntiFLipper, LFighter and Multi-Krum lie within half a point of one another (73.3, 73.2 and 72.9\%). On MNIST, AntiFLipper, Multi-Krum, FLAME and LFighter all reach 98.5--98.7\%, while FedAvg stays near 50\%. Detection separates the methods more clearly than accuracy. AntiFLipper is the only method that finds every attacker without excluding an honest client in all four settings. LFighter does so only with IID data, and Multi-Krum, FLAME and FLTrust exclude honest clients in every setting. The larger accuracy differences appear with skewed data, which the next subsections examine.

\subsection{Honest clients under strong skew}
\begin{figure}[t]
\centering
\includegraphics[width=\columnwidth]{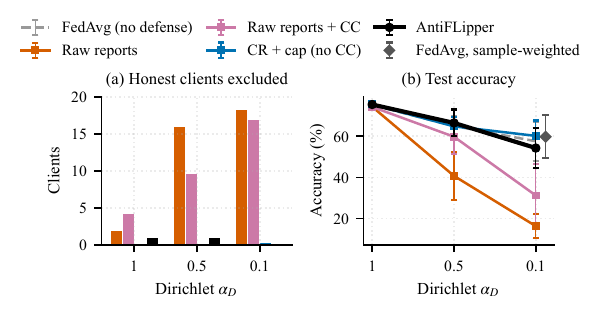}
\caption{Honest clients excluded (a) and test accuracy (b) on CIFAR-10 without attackers, as data become more skewed (smaller $\alpha_D$). Sample-weighted FedAvg was run at $\alpha_D{=}0.1$ only.}
\label{fig:het}
\end{figure}

\begin{figure}[t]
\centering
\includegraphics[width=\columnwidth]{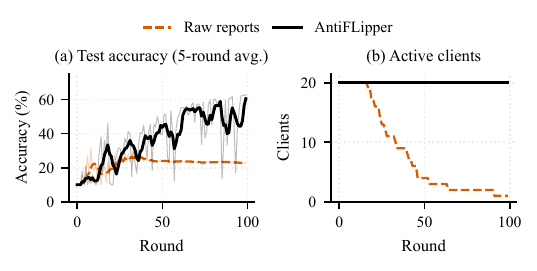}
\caption{Test accuracy (a) and active clients (b) per round in one run without attackers at $\alpha_D{=}0.1$ (seed \trajSeed, the middle of our three seeds by number, not by outcome). Raw reports exclude honest clients in a cascade, whereas AntiFLipper excludes none.}
\label{fig:traj}
\end{figure}

Without attackers every exclusion is a false positive (Fig.~\ref{fig:het}). Raw reports exclude \cleanHonMidRaw{} of 20 honest clients at $\alpha_D{=}0.5$ and \cleanHonRaw{} at $\alpha_D{=}0.1$, where accuracy collapses to \cleanAccRaw\%. Fig.~\ref{fig:traj} shows the resulting cascade, in which exclusions accumulate round after round (Section~\ref{sec:analysis} explains why). With the residual, exclusions stay near zero at every level of skew, and at $\alpha_D{=}0.5$ accuracy rises from \cleanAccMidRaw\% to \cleanAccMid\%.

The residual removes the penalty of raw feedback but does not make strongly skewed training more accurate than training without a defense. At $\alpha_D{=}0.1$, AntiFLipper averages \cleanAcc$\pm$\cleanAccStd\%, equal-weight FedAvg \fedCleanAcc$\pm$\fedCleanStd\% and sample-weighted FedAvg \fedWCleanAcc$\pm$\fedWCleanStd\%, which is ahead of AntiFLipper in each seed. As CC marked only \ccOverrideShare\% of client-rounds and no honest client was excluded, the gap stems from the trust weights, which shift the trajectory considerably at this skew. Without attackers, AntiFLipper thus costs a few points of accuracy under strong skew, in exchange for the protection measured next.

\subsection{Attackers under strong skew}
\begin{table}[t]
\centering
\caption{Strong skew with attackers (CIFAR-10, $\alpha_D{=}0.1$, 40\% attackers, three seeds).}
\label{tab:skew}
\footnotesize
\setlength{\tabcolsep}{3.2pt}
\begin{tabular}{@{}l cccc@{}}
\toprule
\textbf{Method} & Acc.\,$\uparrow$ & ASR\,$\downarrow$ & Rec.\,$\uparrow$ & FPR\,$\downarrow$ \\
\midrule
\multicolumn{5}{@{}l}{\textit{Honest reports}} \\
FedAvg & 43.8{\scriptsize$\pm$9.1} & 18.7 & n/a & n/a \\
FedAvg (weighted) & 44.8{\scriptsize$\pm$9.2} & 21.4 & n/a & n/a \\
LFighter & 41.6{\scriptsize$\pm$11.5} & 19.0 & 0.21 & 0.28 \\
FLTrust (reference)$^{\dagger}$ & 39.3{\scriptsize$\pm$6.5} & 16.0 & 0.58 & 0.53 \\
FLTrust (BN-safe)$^{\dagger}$ & 37.8{\scriptsize$\pm$8.3} & 13.3 & 0.42 & 0.17 \\
Raw reports & 13.0{\scriptsize$\pm$2.6} & 15.8 & 0.88 & 0.94 \\
\textbf{AntiFLipper} & \textbf{49.1}{\scriptsize$\pm$11.9} & \textbf{7.5} & \textbf{0.96} & \textbf{0.00} \\
\midrule
\multicolumn{5}{@{}l}{\textit{Faked reports}} \\
CR + cap (no CC) & \textbf{44.5}{\scriptsize$\pm$7.4} & \textbf{17.2} & 0.00 & \textbf{0.00} \\
\textbf{AntiFLipper} & 43.1{\scriptsize$\pm$9.8} & \textbf{17.2} & \textbf{0.04} & 0.06 \\
\bottomrule
\end{tabular}
\\[2pt]\parbox{\linewidth}{\scriptsize Weighted: FedAvg weighted by client dataset size~\cite{fedavg-pmlr-v54-mcmahan17a}. Faking: attackers report statistics on their original labels. FedAvg, LFighter and FLTrust ignore reports, so faking does not affect them. $^{\dagger}$Uses 100 clean root samples at the server.}
\end{table}

Table~\ref{tab:skew} adds 40\% attackers at $\alpha_D{=}0.1$. AntiFLipper reaches \skewAcc\%, against \skewAccFedAvg\% for FedAvg, \fedWSkewAcc\% for sample-weighted FedAvg, \skewAccLFighter\% for LFighter and \skewAccFLTrust\% for FLTrust. Seed by seed it beats FedAvg in \skewWinsFed{} of 3 seeds and sample-weighted FedAvg in \skewWinsFedW, and its ASR (\skewASR\%, against \skewASRFedAvg\% and \fedWSkewASR\%) is the lowest in the upper part of Table~\ref{tab:skew} in each of the \skewASRBest{} seeds. Weighting by size raises the ASR, because large attackers gain influence. AntiFLipper finds \skewRecall{} of the attackers without excluding an honest client, whereas raw reports flag \skewFPRRaw{} of the honest clients. Detection is slower and not always complete. All attackers were excluded in \skewFull{} seeds, by round \skewDetHits{} on average, and in the third seed \skewMissed{} attacker remained. The probability that a random unmarked attacker has a lower residual than a random honest client (the AUC, averaged over the first ten rounds) is \aucHighMin--\aucHighMax{} per seed at $\alpha_D{=}1$ but only \aucLowMin--\aucLowMax{} at $\alpha_D{=}0.1$, consistent with the estimation bias of Section~\ref{sec:analysis}.

\subsection{Attackers that try to stay hidden}
\label{sec:hidden}
\begin{table*}[t]
\centering
\caption{Partial flips and partial participation (CIFAR-10, $\alpha_D{=}1$, 40\% attackers, three seeds).}
\label{tab:partial}
\footnotesize
\setlength{\tabcolsep}{3pt}
\begin{tabular}{@{}l cccc cccc cccc@{}}
\toprule
& \multicolumn{4}{c}{\textbf{Partial flip, 50\% of labels}} & \multicolumn{4}{c}{\textbf{Partial flip, 25\% of labels}} & \multicolumn{4}{c}{\textbf{Full flip, 50\% participation}} \\
\cmidrule(lr){2-5}\cmidrule(lr){6-9}\cmidrule(l){10-13}
\textbf{Method} & Acc.\,$\uparrow$ & ASR\,$\downarrow$ & Rec.\,$\uparrow$ & FPR\,$\downarrow$ & Acc.\,$\uparrow$ & ASR\,$\downarrow$ & Rec.\,$\uparrow$ & FPR\,$\downarrow$ & Acc.\,$\uparrow$ & ASR\,$\downarrow$ & Rec.$^{*}$\,$\uparrow$ & FPR$^{*}$\,$\downarrow$ \\
\midrule
FedAvg & \textbf{71.9}{\scriptsize$\pm$0.3} & 5.0 & n/a & n/a & \textbf{73.6}{\scriptsize$\pm$0.1} & 3.7 & n/a & n/a & 46.1{\scriptsize$\pm$3.0} & 30.6 & n/a & n/a \\
LFighter & 70.4{\scriptsize$\pm$0.9} & 4.2 & 0.46 & 0.08 & 72.3{\scriptsize$\pm$1.5} & 3.7 & 0.17 & 0.14 & 63.6{\scriptsize$\pm$6.5} & 10.0 & 0.84 & 0.13 \\
FLTrust (BN-safe)$^{\dagger}$ & 70.5{\scriptsize$\pm$0.4} & 4.5 & 0.67 & 0.31 & 66.1{\scriptsize$\pm$8.5} & 4.3 & 0.50 & 0.19 & 62.1{\scriptsize$\pm$2.6} & 11.1 & 0.50 & 0.15 \\
\midrule
\textbf{AntiFLipper} & 71.3{\scriptsize$\pm$0.4} & \textbf{3.2} & \textbf{1.00} & \textbf{0.00} & 72.1{\scriptsize$\pm$0.5} & \textbf{3.2} & \textbf{1.00} & \textbf{0.00} & \textbf{71.0}{\scriptsize$\pm$1.0} & \textbf{3.3} & \textbf{1.00} & \textbf{0.00} \\
\bottomrule
\end{tabular}
\\[2pt]\parbox{\linewidth}{\scriptsize Partial flip: share of each attacker's samples relabelled in every round. Participation: a random half of the active clients is sampled in every round. $^{\dagger}$Uses 100 clean root samples at the server. $^{*}$AntiFLipper: clients excluded in the final round. Baselines: share of sampled attackers (honest clients) given zero weight in the last 10 rounds.}
\end{table*}

\textbf{Partial flips.} An attacker that flips only part of its labels reports a higher accuracy and moves the model less. At both tested fractions AntiFLipper still excludes every attacker without false positives, by round \pfDetHalf{} at 50\% and \pfDetQuarter{} at 25\% (Table~\ref{tab:partial}). LFighter finds only \pfRecLFighterHalf{} and \pfRecLFighterQuarter{} of the attackers and the BN-safe FLTrust \pfRecFLTrustHalf{} and \pfRecFLTrustQuarter, both with false positives. These attacks are weak, however, and FedAvg still reaches \pfAccFedAvgHalf\% and \pfAccFedAvgQuarter\%, slightly above AntiFLipper (\pfAccHalf\% and \pfAccQuarter\%), which loses the correctly labelled part of the excluded clients' data. Partial flips thus show that detection survives a weaker signal, not that it improves accuracy.

\textbf{Report faking under moderate skew.} AntiFLipper detects every faker (Table~\ref{tab:ablation}), by round \fakeDet{} on average, with no false positives. Raw reports find \fakeRecallRaw{} of them and the residual alone \fakeRecallCROnly. Report-only variants find some fakers because the model has trained on their flipped samples, a leakage discussed in Section~\ref{sec:analysis}. The residual of faked reports therefore still ranks attackers below honest clients (AUC \fakeAucResHighMin--\fakeAucResHighMax), but too weakly to exclude them all, and the consistency check catches the rest.

\textbf{Report faking under strong skew.} Here faking succeeds (lower part of Table~\ref{tab:skew}). AntiFLipper finds \fsRecall{} of the attackers (\fsFound{} of \fsTotal{} over three seeds), and its accuracy (\fsAcc\%) is no better than that of FedAvg (\fsAccFedAvg\%). The logs show why. The raw faked reports still rank attackers below honest clients (AUC \fakeAucRawLowMin--\fakeAucRawLowMax), but the residual reduces this to \fakeAucResLowMin--\fakeAucResLowMax. Under strong skew each class is concentrated on a few clients, so for classes held mostly by attackers, $\hat A_k$ consists largely of their own reports and absorbs their deficit, the bias of Section~\ref{sec:analysis}. The consistency check alone finds \ccSkewRecall{} of the attackers, because honest updates already differ widely. This is the main open case. A reference that leaves out the scored client and a robust per-class estimate are natural next steps.

\subsection{Partial participation}
\label{sec:participation}
In cross-device FL only part of the clients take part in a round. With a random half of the active clients sampled in every round (Table~\ref{tab:partial}, right), AntiFLipper reaches \ppAcc$\pm$\ppAccStd\%, close to its \headAcc\% with full participation, and again excludes every attacker without false positives in \ppFull{} seeds. Detection is slower (round \ppDet{} on average instead of \headDet), because each client collects flags only in the rounds in which it is sampled. LFighter reaches \ppAccLFighter$\pm$\ppAccLFighterStd\%, the BN-safe FLTrust \ppAccFLTrust\% and FedAvg \ppAccFedAvg\%, with ASRs of \ppASRLFighter\%, \ppASRFLTrust\% and \ppASRFedAvg\% against \ppASR\% for AntiFLipper. Baselines judge each round anew, whereas persistent trust lets AntiFLipper use evidence from every round in which a client took part.

\subsection{Ablation}
Table~\ref{tab:ablation} attributes these results to components. In the main setting all feedback variants perform similarly, and CC alone, with the same $z$ and cap, is as accurate as AntiFLipper (\ccHeadAcc\%). The harder settings separate them. Raw reports fail without attackers under strong skew and against report faking, and the cap alone only limits the damage (\cleanHonBudget{} honest clients excluded instead of \cleanHonRaw). The residual prevents these exclusions but misses fakers. CC alone catches every faker but finds only \ccSkewRecall{} of the attackers under strong skew. Each signal thus covers the failure of the other, and only the combination does well in all three settings. The combination is not best everywhere, since without attackers under strong skew the residual alone is more accurate.

\subsection{Share of malicious clients}
\begin{figure}[t]
\centering
\includegraphics[width=\columnwidth]{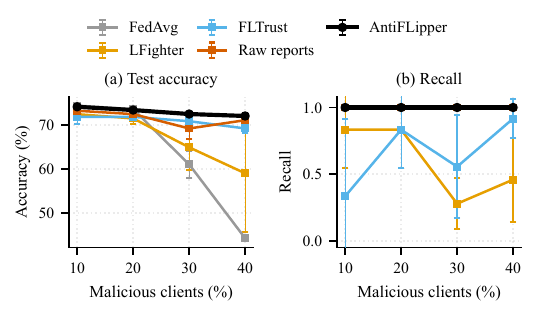}
\caption{Test accuracy (a) and recall (b) on CIFAR-10 non-IID ($\alpha_D{=}1$) versus the share of malicious clients.}
\label{fig:ratio}
\end{figure}
From 10\% to 40\% attackers (Fig.~\ref{fig:ratio}), AntiFLipper keeps a recall of 1.0 with an FPR of at most \ratioFPRmax{} (one or two honest clients in some seeds). At 10\% it matches undefended FedAvg (\ratioAccLow\% vs.\ \ratioAccFedAvgLow\%), so the defense costs nothing when the attack is weak. At 30\% it reaches \ratioAccThirty\%, while FedAvg drops to \ratioAccFedAvgThirty\%, LFighter to \ratioAccLFighterThirty\% and FLTrust to \ratioAccFLTrustThirty\%.

\subsection{Cost and sensitivity}
\begin{table}[t]
\centering
\caption{Cost per round (CIFAR-10, ResNet-18, 20 clients).}
\label{tab:cost}
\footnotesize
\setlength{\tabcolsep}{3.2pt}
\begin{tabular}{@{}l ccccc@{}}
\toprule
& \textbf{Server} & \multicolumn{2}{c}{\textbf{Server time (ms)}} & \textbf{Client} & \textbf{Report} \\
\cmidrule(lr){3-4}
\textbf{Method} & \textbf{data} & matched & in training & \textbf{overhead} & \textbf{payload} \\
\midrule
FedAvg & \ding{55} & 183 & 34 & 0 & 0 \\
Median & \ding{55} & 1599 & 356 & 0 & 0 \\
Multi-Krum & \ding{55} & 4109 & 875 & 0 & 0 \\
FLAME & \ding{55} & 3465 & 1058 & 0 & 0 \\
LFighter & \ding{55} & 214 & 42 & 0 & 0 \\
FLTrust (reference)$^{\dagger}$ & \ding{51} & 1269 & 352 + 115 & 0 & 0 \\
\midrule
\textbf{AntiFLipper} & \ding{55} & 173 & 30 & 6.8\% & 84\,B \\
\bottomrule
\end{tabular}
\\[2pt]\parbox{\linewidth}{\scriptsize Matched: identical 20 updates on one CPU (Intel i7-8700, 6 threads, PyTorch 2.8), mean of 5 timed iterations after one warm-up, with standard deviation 8\,ms for AntiFLipper and 25\,ms for FedAvg. In training: measured per round on an RTX 3090 GPU, where excluded clients are not averaged. $^{\dagger}$Uses 100 clean root samples at the server. Its in-training time is aggregation + root training.}
\end{table}

In the matched benchmark (Table~\ref{tab:cost}), AntiFLipper's server work takes \matchedOurs\,ms, comparable to FedAvg (\matchedFedAvg\,ms), against \matchedFLTrust\,ms for FLTrust without root training, \matchedKrum\,ms for Multi-Krum and \matchedFLAME\,ms for FLAME. During training we observed \defMs\,ms per round, versus \defMsFedAvg\,ms for FedAvg, which keeps averaging all models after AntiFLipper has excluded the attackers. The client evaluation adds \evalOverhead\% to local training. With 32-bit model values, the 84-byte report adds approximately $0.096\%$ to the $87{,}360$-byte MNIST CNN upload and $0.00019\%$ to the $44{,}764{,}968$-byte CIFAR-10 ResNet-18 upload.

For sensitivity, we varied the flag limit $f_{\max}\in\{3,6,9\}$ and the cut-off $z\in\{2.5,3,3.5\}$ one at a time on MNIST non-IID with held-out seeds. With 40\% attackers, accuracy stayed within \sensAccMin--\sensAccMax\% and recall at \sensRecallMin{} in every setting. Only the FPR without attackers reacted to the most aggressive settings ($f_{\max}{=}3$, $z{=}2.5$), reaching \sensFPRmax{} in a single run, which motivates the defaults. MNIST is an easy task, so this shows robustness to the parameters rather than to the data.

\section{Analysis}
\label{sec:analysis}
This section explains the behavior observed in Section~\ref{sec:results}. Let $A_k$ be the accuracy of $M_t$ on class $k$ and $C_{kl}$ the rate at which it predicts class $l$ for inputs of class $k$. We assume that these rates do not depend on which client a sample comes from.

\begin{proposition}[Raw feedback penalizes label mix]\label{prop:mix}
For an honest client, $\mathbb{E}[a_j]=\sum_k\pi_{jk}A_k$, so its deviation from the population mean is $\sum_k(\pi_{jk}-\bar\pi_k)A_k$, where $\bar\pi_k$ is the population share of class $k$.
\end{proposition}

The deviation is negative whenever a client holds more than its share of hard classes. It vanishes under IID data and is large when each client holds few classes, which is the first failure mode.

\begin{proposition}[Residual with exact per-class accuracies]\label{prop:cr}
If $\hat A_k=A_k$ for all $k$, an honest client has $\mathbb{E}[r_j]=0$ for every label mix, while a flipper that relabels all its samples and reports honestly has $\mathbb{E}[r_j]=-\sum_k\pi_{jk}\,\big(A_{\sigma(k)}-C_{k\sigma(k)}\big)$, where $\pi_{jk}$ here denotes the share of its samples whose original class is $k$.
\end{proposition}
\emph{Proof.} The first claim follows from~(\ref{eq:cr}) and Prop.~\ref{prop:mix}. A flipper's samples of true class $k$ carry label $\sigma(k)$. They are counted under $\sigma(k)$ and are correct only when $M_t$ predicts $\sigma(k)$, at rate $C_{k\sigma(k)}$, whereas the expectation uses $A_{\sigma(k)}$. If a random fraction $q$ is relabelled, as in our partial flips, the deficit scales by $q$, so detection takes longer (Section~\ref{sec:hidden}).

\textbf{Estimation bias.} In practice $\hat A_k$ is estimated in the presence of flippers. Let $w_k$ be their trust-weighted share of the samples \emph{labelled} $k$. Then $\hat A_k$ is pulled towards their low class-$k$ accuracy by a factor of up to $w_k$, which exceeds their overall trust share when they hold many samples of class $k$. Because each client's own report enters $\hat A_k$, a flipper that holds most of a class pulls the reference towards itself and hides part of its own deficit. Separation therefore requires small $w_k$, which trust weighting promotes but strong skew can violate.

\begin{proposition}[Report faking defeats report-only detection]\label{prop:faking}
Let a flipper report the statistics of $M_t$ on its original labels, and let $M_t$ not depend on the labels it trains with. Then its report has the same distribution as that of an honest client with identically distributed data, and no detector that reads only reports can separate the two better than chance.
\end{proposition}
\emph{Proof.} The faked report is the same function of $(M_t,x,y)$ as an honest report, and by assumption $M_t$ carries no trace of the flipped labels.

The assumption holds in the first round. Later, $M_t$ has been trained on the flipper's own flipped samples, so a faked report may still fall below an honest one, as measured in Section~\ref{sec:hidden}. A defense cannot rely on it, so it must also observe what a client \emph{did}. Relabelling class $k$ as $\sigma(k)$ pushes the classifier row of $\sigma(k)$ towards the features of class $k$ and vice versa, which we expect to oppose the last-layer change of honest clients that hold the same classes. When fewer than half of the sampled updates are malicious, every coordinate of $\tilde\delta$ lies between honest values~\cite{yin2018byzantine}. This bounds the reference but does not guarantee separation, since honest similarities also spread under heterogeneity, so we use a conservative cut-off ($z{=}3$) and measure the false-positive rate.

\textbf{Exclusion bound and convergence.} The cap keeps at least $0.6N$ clients, so no sequence of false positives can empty the federation. We do not claim a convergence guarantee, because even with fixed membership the trust weights in M6 change over time, so standard FedAvg analyses~\cite{li2020convergence} do not apply directly. We report trajectories instead (Fig.~\ref{fig:traj}).

\textbf{Complexity.} The report adds one forward pass over $D_j$ to $E$ training epochs, a client cost of roughly $1/(3E)$. On the server, M2 costs $O(n_tK)$ and M4 $O(n_t d_L\log n_t)$ for the coordinate-wise median over a last layer of $d_L$ parameters (5{,}130 for ResNet-18), whereas Multi-Krum costs $O(n_t^2W)$ for a $W$-parameter model.

\textbf{Privacy.} Per-class counts reveal a client's label histogram. Since the server needs only the pooled sums in~(\ref{eq:cr}), secure aggregation of these sums would let each client compute $r_j$ locally from the broadcast $\hat A$ and send one scalar, a variant we have not implemented.

\section{Limitations}
\label{sec:limits}
AntiFLipper does not solve report faking under strong skew (\fsFound{} of \fsTotal{} attackers detected), and an attacker that also aligns its last-layer update with the median could evade CC under moderate skew. Colluding attackers were not evaluated. Under strong skew with honest reporting, AntiFLipper detected 23 of \fsTotal{} attackers and needed more rounds than under moderate skew. Without attackers under strong skew, it costs a few points of accuracy against FedAvg (\cleanAcc\% vs.\ \fedCleanAcc\%), and combining its trust with sample weighting is untested. Our analysis does not establish convergence under time-varying trust weights, and per-class counts reveal label histograms unless the secure-aggregation variant is deployed. Other domains, asynchronous updates, other participation rates and data-dependent dropout remain untested. Three seeds support means and standard deviations, not fine-grained significance claims.

\section{Conclusion}
\label{sec:conclusion}
Accuracy feedback is a cheap, data-free signal against label flipping, but used raw it mistakes skewed honest clients for attackers and can be faked. AntiFLipper judges each report against the client's own label mix and checks the update itself, at a server cost comparable to FedAvg and without server data. On moderately skewed CIFAR-10 with full flips it was more accurate than all six baselines, and on MNIST it matched the best of them. It was the only tested method to detect every attacker without false positives on both data sets, IID and non-IID, and on moderately skewed CIFAR-10 it kept doing so under partial flips, faked reports and 50\% participation. Under strong skew it excluded no honest client without attackers, against \cleanHonRaw{} of 20 with raw reports, and with attackers it gave a lower ASR than every baseline in every seed. Report faking under strong skew remains open (\fsFound{} of \fsTotal{} attackers detected), because there the label-mix correction that protects honest clients also hides the attackers. Resolving this tension is the main direction for future work.

\bibliographystyle{IEEEtran}
\bibliography{IEEEabrv,bibliography_updated}
\end{document}